# Lorentzian Switching Dynamics in HZO-based FeMEMS Synapses for Neuromorphic Weight Storage

Shubham Jadhav[1,a], Kaustav Roy[1], Luis Amaro[1], Thejas Basavarajappa[1], Madhav Ramesh[1], Debdeep Jena[1,2], Huili (Grace) Xing[1,2], and Amit Lal[1,a]

[1]School of Electrical and Computer Engineering, Cornell University, Ithaca, New York 14853, USA
[2]Department of Materials Science and Engineering, Cornell University, Ithaca, New York 14853, USA
[a]Author to whom correspondence should be addressed: saj96@cornell.edu, amit.lal@cornell.edu

**Abstract:**

Neuromorphic computing demands synaptic elements that can store and update weights with high precision while being read non-destructively. Conventional ferroelectric synapses store weights in remnant polarization states and might require destructive electrical readout, limiting endurance and reliability. We demonstrate a ferroelectric MEMS (FeMEMS) based synapse in which analog weights are stored in the piezoelectric coefficient $d_{31,eff}$ of a released $Hf_{0.5}Zr_{0.5}O_2$ (HZO) MEMS unimorph. Partial switching of ferroelectric domains modulates $d_{31,eff}$, and a low-amplitude mechanical drive reads out the weight without read disturb in the device yielding more than 7-bit of programming levels. The mechanical switching distribution function follows a Lorentzian distribution as a logarithmic function of partial poling voltage ($V_p$) consistent with nucleation-limited switching (NLS), and the median threshold extracted from electromechanical data obeys a Merz-type field–time law with a dimensionless exponent $\alpha = 3.62$. These relationships establish a quantitative link between mechanical weights and electrical switching kinetics. This mechanically read synapse avoids depolarization and charge-injection effects, provides bipolar weights (well suited for excitatory and inhibitory synapses), directly reveals partial domain populations, and offers a robust, energy-efficient route toward high-bit neuromorphic hardware.

Neuromorphic computing seeks to emulate the analogue weight storage and low-power operation of biological synapses[1–11]. Ferroelectric hafnium-zirconium oxide (HZO) is highly attractive because it maintains a stable ferroelectric phase even at ultra-thin dimensions (below ~20 nm), which enables (a) reduced defect density, (b) lower read/write voltages, and (c) a smaller device footprint for higher integration density. Moreover, HZO thin films demonstrate strong reliability, endurance, and retention, while being CMOS-compatible and deployable in commercial fabrication processes[12–18]. Conventional ferroelectric synapses store the weight in the remnant polarization ($P_r$) states remnant of a ferroelectric field effect transistor (FeFET)[19–23], a ferroelectric capacitor (FeCAP)[24,25], and a ferroelectric tunnel junction (FTJ)[26–29]. Partial polarization reversal of the ferroelectric layer enables quasi-continuous (or multiple discrete) modulation of conductance, resistance or capacitance, and has thus been used to achieve multiple distinct weight levels for neuromorphic (synapse-like) operation. However, electrical read-out of $P_r$ might perturb the ferroelectric state thereby introducing depolarization fields and charge-injection currents[30] and requires large programming voltages. These non-idealities limit endurance and make it difficult to achieve reliable weight updates across the range of operation[31–35].

An alternative route is to store weights in the volume-integrated piezoelectric coefficient $d_{31,eff}$, rather than in the interface-influenced remnant polarization that governs FeFETs and other conventional ferroelectric devices. All ferroelectrics are also piezoelectric; partial switching of domains alters $d_{31,eff}$, and a mechanical read-out can thus provide a non-destructive measure of the switched fraction unlike the conventional ferroelectric devices. Building on this concept, we recently demonstrated that partial ferroelectric switching in a released HZO unimorph beam allows programmable control of $d_{31,eff}$ and, consequently, of the beam displacement under a small mechanical drive[36]. The displacement is proportional to the product of the programmed $d_{31,eff}$ and the input voltage, enabling direct multiply operations. Such multipliers when connected in parallel enables the accumulate operation thereby enabling multiply-and-accumulate operations. In follow-up work we employed scandium-alloyed AlScN films and achieved similar control of the piezoelectric coefficient in a FeMEMS multiplier with zero stand-by leakage and linear displacement versus input voltage[37].

The underlying physics of domain switching is the same for $P_r$ and $d_{31}$ — both originate from the ensemble of switched and unswitched regions. However, FeMEMS synapses access the volume-integrated piezoelectric coefficient, thereby enabling a quantitative handle on a wider range of partial polarization fractions, resulting in finer synaptic weights beyond what is possible from $P_r$ alone. Furthermore, mechanical read-out is non-destructive in nature: a small sinusoidal drive excites out-of-plane motion without read-induced switching of the film. Because both positive and negative $d_{31,eff}$ values are accessible, the mechanical weight can naturally assume positive or negative sign, which is useful for implementing excitatory and inhibitory synapses. These advantages of FeMEMS synapses over the conventional ferroelectric synapses provides a route to develop a neuromorphic hardware with direct, reliable, and non-destructive read-out.

The distribution of local switching thresholds in disordered ferroelectric films is broad; nucleation-limited switching (NLS) models capture this behaviour by treating the film as an ensemble of independent switching units with a statistical distribution of switching times. Experiments on HZO and PZT films reveal that the cumulative switching fraction versus electric field is well described by a Lorentzian (Cauchy) distribution of thresholds[38], and the median switching time obeys an empirical Merz law[39], in which the switching time scales exponentially with the reciprocal of the electric field. These insights, developed originally for electrical switching, also apply to the mechanical $d_{31}$ weights: by mapping the measured displacement to the fraction of switched domains, we show below that the weight transfer function follows a Lorentzian distribution in the common-logarithm ($\log_{10}$) of the programming voltage (see Supplementary Equation (S7)) and that the median switching threshold obeys a Merz-type field–time scaling. Knowledge of the underlying distribution allows us to predict and calibrate the number of attainable weight levels with high precision.

Thus, in this paper we demonstrate a FeMEMS based neuromorphic synapse that writes analog weights in the volume-integrated piezoelectric coefficient $d_{31,eff}$ of a released HZO unimorph beam and offers an added advantage of non-destructive readout. We begin by describing the device structure, fabrication and programming/readout scheme, thereby presenting the experimental results, including the correlation between electrical polarization and mechanical displacement, repeatability, pulse-width dependence and extraction of Lorentzian parameters. Subsequently we develop a quantitative model based on NLS and Merz law to interpret the electromechanical data and assess the precision of weight storage Finally, we determine the number of distinct programming levels/weights attainable using our device to be $\sim 200$ thereby discussing the implications for neuromorphic hardware and outlining potential pathways to high-bit mechanical synapses.

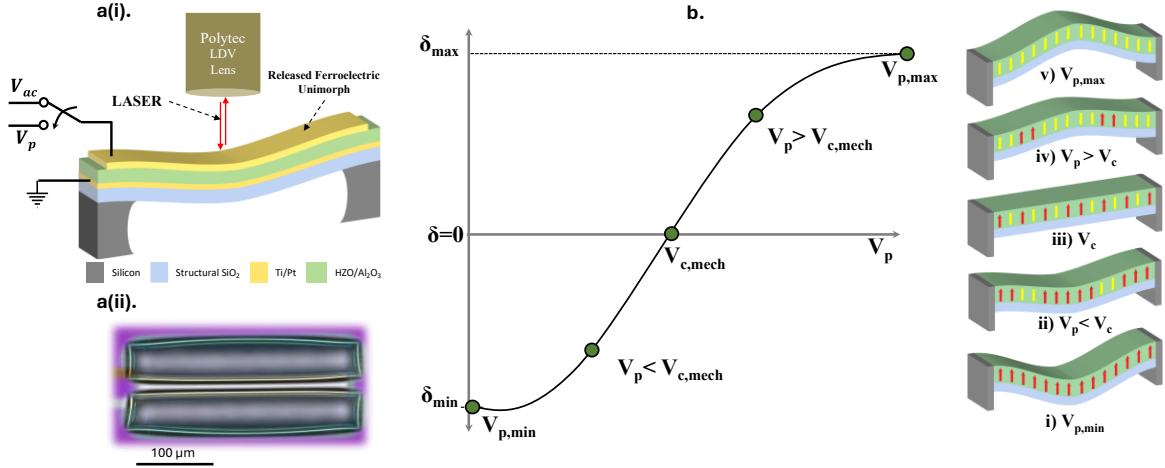

**Figure 1a(i).** Measurement schematic for a released ferroelectric MEMS unimorph. The $HZO/Al_2O_3/Ti/Pt/SiO_2/Si$ stack is partially poled using a triangular voltage pulse of amplitude $V_p$ and width $t_p$. After poling, a small-signal sinusoidal drive $V_{ac}$ ($V_{ac} \ll V_p$) actuates the beam, and the out-of-plane tip displacement $\delta$ is recorded with a Polytec MSA-400 laser Doppler vibrometer (LDV). **1a(ii).** Optical micrograph of the HZO based FeMEMS unimorph used for the experiments. **Fig. 1b.** Maximum beam displacement versus poling amplitude demonstrating reversal and tuning of the effective piezoelectric coefficient $d_{(31,eff)}$. Starting from the reset state ($\delta = \delta_{min}$ at $V_{(p,min)}$), increasing $V_p$ progressively switches domains, moving the device toward $\delta_{max}$ at $V_{(p,max)}$. The curve crosses $\delta = 0$ at the mechanical coercive voltage $V_{(c,mech)}$, where up- and down-polarized domain fractions balance and $d_{(31,eff)} \approx 0$. Insets **(i–v)** illustrate representative domain configurations along the sweep.

The neuromorphic weight element is a clamped–clamped unimorph beam engineered so that its out-of-plane motion reflects the effective piezoelectric coefficient. Figure 1a(i) illustrates the device geometry: from bottom to top, the layer stack comprises a Si substrate, a 200-nm structural $SiO_2$ layer, a Ti/Pt bottom electrode (5/50 nm), a 17-nm HZO ferroelectric film, a 3-nm $Al_2O_3$ cap and a Ti/Pt top electrode (5/50 nm). The structural oxide forms the elastic layer. the ferroelectric $HZO/Al_2O_3$ stack stores the weight and provides piezoelectric actuation, and the Pt electrodes deliver the programming and readout voltages. The thicknesses are chosen so that the neutral axis lies within the ferroelectric, maximizing the induced strain for a given electric field. Test beams have a length $L \approx 300\ \mu m$ and width $W \approx 24\ \mu m$, with the ferroelectric thickness thin enough to ensure single-phase orthorhombic HZO. Figure 1a(ii) illustrates an optical micrograph of the HZO based FeMEMS unimorph used for the experiments. Figure 1b plots the maximum beam displacement $\delta$ versus $V_p$ at fixed partial poling time ($t_p$), showing reversible and continuous tuning of the $d_{31,eff}$; the curve crosses zero at the mechanical coercive voltage ($V_{c,mech}$) where up- and down-polarized domain fractions balance each other.

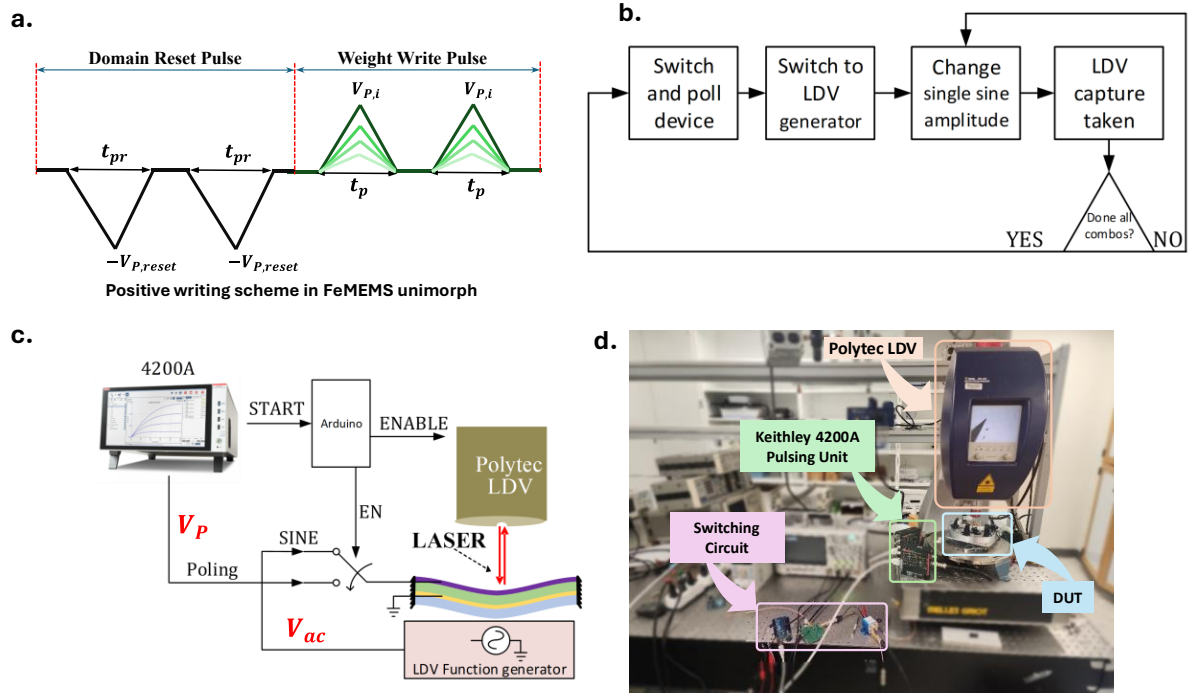

**Figure 2a.** Pulse protocol used to write the weight (stored in the effective piezoelectric coefficient ($d_{31,\text{eff}}$)) Each step begins with two negative reset pulses of amplitude $V_{p,\text{reset}}$ and width $t_{pr}$ to reproducibly initialize the state, followed by two triangular write pulses with peak $V_{p,i}$ and width $t_p$. This sequence sets the domain fraction that determines $d_{31,\text{eff}}$. The sequence is repeated for successive $V_{p,i}$ values to program different $d_{31,\text{eff}}$ (weights). **2b.** Automated measurement flow for weight writing and readout. For each target ($V_{p,i}, t_p$): apply the switching pulse, route the top electrode to the LDV generator, set the single-tone actuation $V_{ac}$, acquire the beam displacement with the LDV, then advance to $V_{p,i+1}$. The loop repeats until all ($V_{p,i}, t_p$) combinations are measured. **2c.** Block diagram of the fully automated setup. A Keithley 4200A pulsing unit applies poling pulses $V_p$ to the top electrode while the bottom electrode is grounded. An Arduino-controlled switch then connects the top electrode to the Polytec LDV's function generator to apply a small-signal sinusoid $V_{ac}$ for readout. The LDV records beam displacement amplitude and phase. **2d.** Photograph of the bench setup showing the Polytec LDV head, Keithley 4200A pulsing unit, Arduino-based switching circuit, and the device under test (DUT).

The neuromorphic weight element was fabricated using a process flow adapted from our previous work[36]. A 200-nm $SiO_2$ layer is deposited on a polished Si wafer by plasma-enhanced chemical vapour deposition. A Ti/Pt bottom electrode is sputtered and patterned lithographically. The ferroelectric HZO and $Al_2O_3$ films are deposited by atomic-layer deposition and annealed at 400 °C for 1 min in nitrogen to crystallize ferroelectric HZO. A Ti/Pt top electrode is sputtered and patterned. The ferroelectric stack and underlying oxide are etched to define the beam and open release windows. Finally, $XeF_2$ isotropic dry etching removes the underlying Si to fully release the clamped–clamped bridge.

Weight programming is accomplished using a sequence of triangular voltage pulses that control the fraction of switched domains. Each programming cycle begins with a reset comprising two negative triangular pulses of peak voltage $V_{p,reset}$ and duration $t_{pr}$, which fully poles the film into the up-polarized state. Following the reset, a write step applies two identical triangular pulses with peak $V_p$ and width $t_p$. These pulses partially switch the domains in down direction, setting an ensemble-averaged switched fraction $S(V_p, t_p)$. Reversing the polarity inverts the writing scheme. Immediately after the write, the top electrode is connected to a function generator via a microcontroller and a small-signal sinusoid of amplitude $V_{ac} \ll V_p$ is applied (Figure 2a–d). A Polytec laser Doppler vibrometer measures. the out-of-plane displacement $\delta(V_p, t_p)$ at a fixed frequency of the beam. Because the displacement is linear in both $d_{31,eff}$ and $V_{ac}$, we can write

$$(1)\ \delta(V_p, t_p) = K_{geom}\, V_{ac}\, d_{31,eff}(V_p, t_p),$$

where $K_{geom}$ depends on beam geometry and mode shape. For convenience the measured displacement can also be expressed in terms of two calibration constants, $\delta_{min}$ and $\delta_{max}$, determined from a full reset and a full pole:

$$(2)\ \delta(V_p, t_p) = \delta_{min} + A * S(V_p, t_p)$$

where, $A = K_{geom} V_{ac}(\delta_{max} - \delta_{min})$, $\delta_{min}$ and $\delta_{max}$ are the displacements of the fully reset and fully poled states, respectively.

Here $S(V_p, t_p)$ represents the ensemble-averaged switched fraction of domains. The mechanical coercive voltage $V_{c,mech}$ is defined as the poling amplitude at which $\delta = 0$; at this point up- and down-polarized domain fractions balance so $d_{31,eff} \approx 0$. Because the small-signal read does not disturb the polarization state, multiple reads can be performed without drift. The calibration constants are measured once per device, allowing the system to map any intermediate displacement directly to a unique weight level.

To establish that weights encoded in the effective piezoelectric coefficient mirror the polarization of the ferroelectric, we simultaneously measured the remnant polarization change $\Delta P(V_p, t_p)$ using a Keithley 4200A pulsing unit and the beam displacement $\delta(V_p, t_p)$ using the LDV for a fixed write-pulse width of 500 µs. Figure 3a shows a family of minor hysteresis loops acquired by sweeping the $V_p$ in 5 mV steps; the remnant polarization varies monotonically with $V_p$. Figure 3b overlays $\Delta P$ (red axis) and $\delta$ (black axis) versus $V_p$. Both observables exhibit sigmoidal transitions between a up-polarized state and an down-polarized state. The mechanical curve crosses zero at the $V_{c,mech} = 5.05\ V$, whereas the electrical midpoint at which $\Delta P$ reaches one half of its saturation value defines the electrical coercive voltage $V_{c,elec} = 5.23\ V$. The small offset between $V_{c,mech}$ and $V_{c,elec}$ is consistent with the different observables and the fact that the mechanical read is performed in the presence of a small AC drive. Importantly, both curves can be mapped onto one another by a linear transformation, validating that the displacement is proportional to the switched domain fraction and therefore encodes the same physics as $\Delta P$.

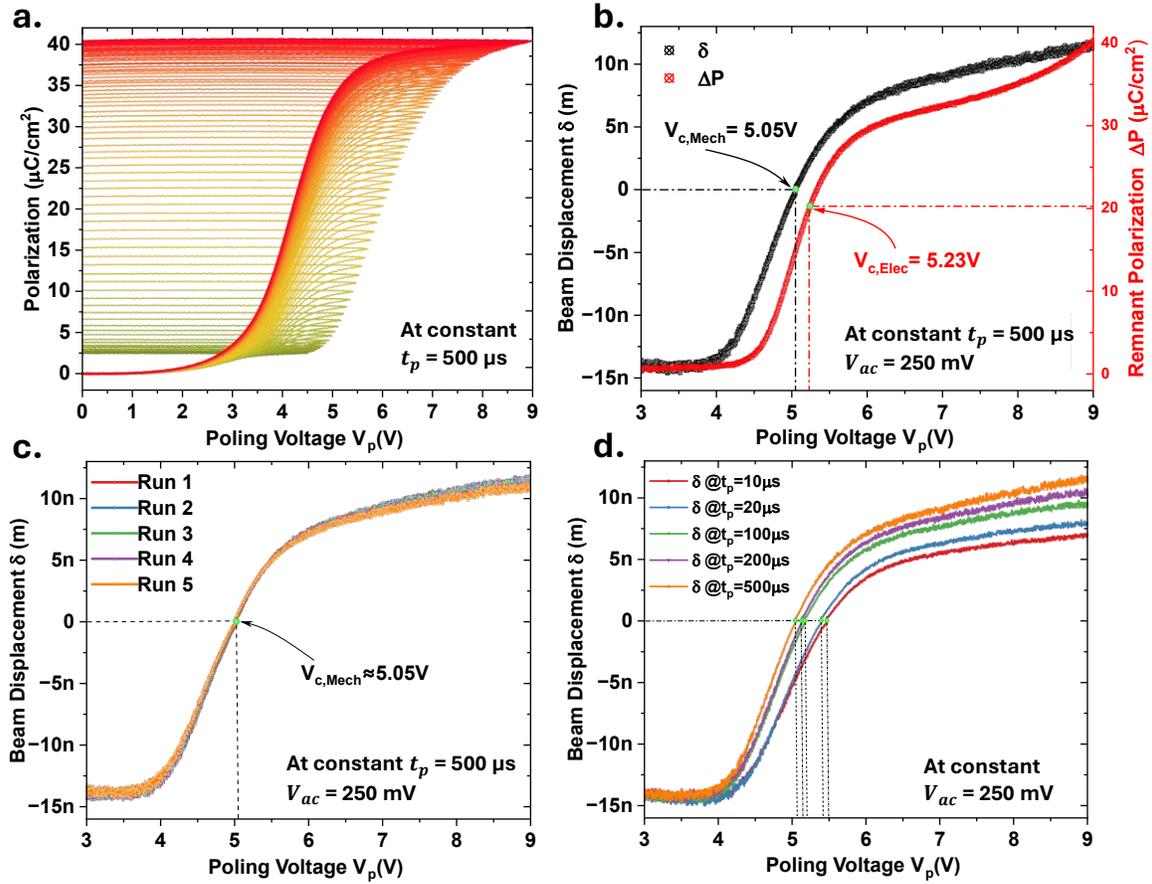

**Figure 3a.** Family of minor $P - V$ loops acquired at fixed $t_p = 500\ \mu s$ using triangular write pulses with peak $V_p$ swept from 0.5 to 9 V in 5 mV steps (subset shown for clarity). For each loop, the remanent polarization change $\Delta P$ is extracted at zero field. **3b.** Overlay of beam displacement $\delta$ (black, left axis) and remanent polarization change $\Delta P$ (red, right axis) versus $V_p$ at $t_p = 500\ \mu s$ and $V_{ac} = 250\ mV$. The mechanical coercive voltage $V_{c,\mathrm{mech}}$ (where $\delta = 0$) is 5.05 $V$; the electrical coercive voltage $V_{c,\mathrm{elec}}$ (where $\Delta P = \frac{1}{2}\Delta P_{\mathrm{max}}$) is 5.23 $V$. The small offset between $V_{c,\mathrm{mech}}$ and $V_{c,\mathrm{elec}}$ is consistent with the superposed read drive $V_{ac}$ and differing observables (mechanical vs polarization). **3c.** Repeatability of weight writing. Displacement $\delta(V_p)$ measured over five independent runs at $t_p = 500\ \mu s$, $V_{ac} = 250\ mV$, and 5 mV $V_p$ increments shows overlap within experimental scatter, indicating reproducible programming of $d_{31,\mathrm{eff}}$. **3d.** Displacement $\delta(V_p)$ for multiple pulse widths $t_p \in \{10, 20, 100, 200, 500\}\mu s$ at fixed $V_{ac} = 250\ mV$. Shorter $t_p$ shifts the switching transition to higher $V_p$, lowering $\delta$ at a given $V_p$; longer $t_p$ shifts it left, consistent with Merz-type field–time switching kinetics.

The $\delta(V_p)$ curve exhibits a minor asymmetry between positive and negative displacements. This asymmetry originates from fixed interface charges and built-in electrostatic fields at the HZO/$Al_2O_3$ and electrode interfaces, arising from stack asymmetry that favors one polarization direction. Similar asymmetries have been reported in PZT films and are attributed to local dipole defects that pin domain walls and broaden the switching distribution[38]. Although the asymmetry marginally shifts the weight transfer function, it is fully captured by the calibration constants $\delta_{\min}$ and $\delta_{\max}$; once calibrated, both positive and negative $d_{31,eff}$ values can be accessed deterministically.

The mechanical readout is linear and noise-free, the uncertainty in each level is dominated by statistical fluctuations in domain switching. Repeated programming shows that the standard deviation of $\delta$ at a given $V_p$ is less than 1 nm (see Fig 3c). To probe the switching kinetics, we measured $\delta(V_p)$ for multiple write-pulse widths $t_p$ ranging from 10 μs to 500 μs at a fixed read voltage. Figure 3d shows that shorter pulses shift the transition to higher $V_p$, whereas longer pulses reduce the required poling amplitude. This behaviour reflects the finite switching time of domains: at shorter $t_p$ only domains with the lowest thresholds can switch, so a larger field is needed to reach a given switched fraction; at longer $t_p$ the effective threshold decreases. Similar field–time scaling has been observed in electrical switching experiments on HZO films and is well described by Merz's empirical law, in which the switching time constant scales as $\tau \propto exp(E_a/E)$ with $E = V_p/t_p$ [39](see Supplementary Equation (S3)). Our mechanical measurements thus inherit the same kinetics, reinforcing that partial switching underlies both the electrical $P_r$ and the mechanical $d_{31,eff}$.

Domain switching in polycrystalline ferroelectric films is governed by nucleation-limited switching (NLS) processes limited by local pinning sites. The NLS model expresses the remnant polarization change as an integral over switching times with a distribution $F(\log t_0)$. When $F$ is broad and Lorentzian, the cumulative switching fraction takes the arctangent form used in our fits. The Lorentzian distribution arises physically from random local fields at domain pinning sites: theory and experiments on disordered ferroelectrics show that dipole defects produce a Lorentzian distribution of local fields, and when the domain wall velocity depends exponentially on the inverse field, the resulting distribution of switching times is Lorentzian in $\log t_p$ [33,35]. This mechanism also explains why our $d_{31,eff}$ transfer function is well described by a Lorentzian distribution in $\log V_p$ — the poling amplitude $V_p$ sets the electric field during the write pulse, and longer pulses select regions with lower thresholds, analogous to the time dependence in electrical switching.

The broadness of the $\delta(V_p)$ transition arises from the distribution of local switching thresholds in the ferroelectric film. To quantify this distribution, we fit the $\delta(V_p)$ curves at each $t_p$ to a logistic function derived from a Lorentzian distribution in $\log_{10} V_p$, defined as (see Supplementary Equation (S9)):

$$(3)\ \delta(V_p, t_p) = y_0 + A\left[\frac{1}{2} + \frac{1}{\pi}\arctan\left(\frac{\log_{10}V_p - \mu(t_p)}{w(t_p)}\right)\right]$$

where $y_0 \equiv \delta_{min}$ and $A \equiv K_{geom}V_{ac}(\delta_{max} - \delta_{min})$, $\mu(t_p)$ is the median threshold in log scale and $w(t_p)$ is the half-width at half-maximum. This form, which has been used to analyse domain switching in disordered ferroelectric films[38,40], naturally arises from the nucleation-limited switching model when the distribution of switching times is Lorentzian in $\log_{10}t$. Fits to our experimental data (see Fig 4a) show excellent agreement and yield values of $\mu(t_p)$ and $w(t_p)$ listed in the Supplementary Information. The widths $w$ are nearly independent of $t_p$, indicating a time-invariant spread of local thresholds (see Fig 4b), whereas the medians shift linearly with $\log_{10} t_p$ (see Fig 4c), consistent with Merz scaling given by $\tau \propto \exp(E_a/E)$. The Lorentzian form also predicts that rescaling the axis by $z = (\log_{10} V_p - \mu(t_p))/w(t_p)$ and plotting the normalized displacement $\bar{S} = (\delta - y_0)/A$ versus $z$ collapses all datasets onto the parameter-free single master Cauchy function

$$(4)\ \bar{S}(z) = \frac{1}{2} + \frac{1}{\pi}\arctan z,$$

as shown in Fig. 4d, which we confirm experimentally. Such data collapse demonstrates that the underlying distribution is universal and provides a powerful calibration tool: by knowing $\mu$ and $w$, any desired weight level can be targeted by choosing an appropriate programming voltage and pulse width.

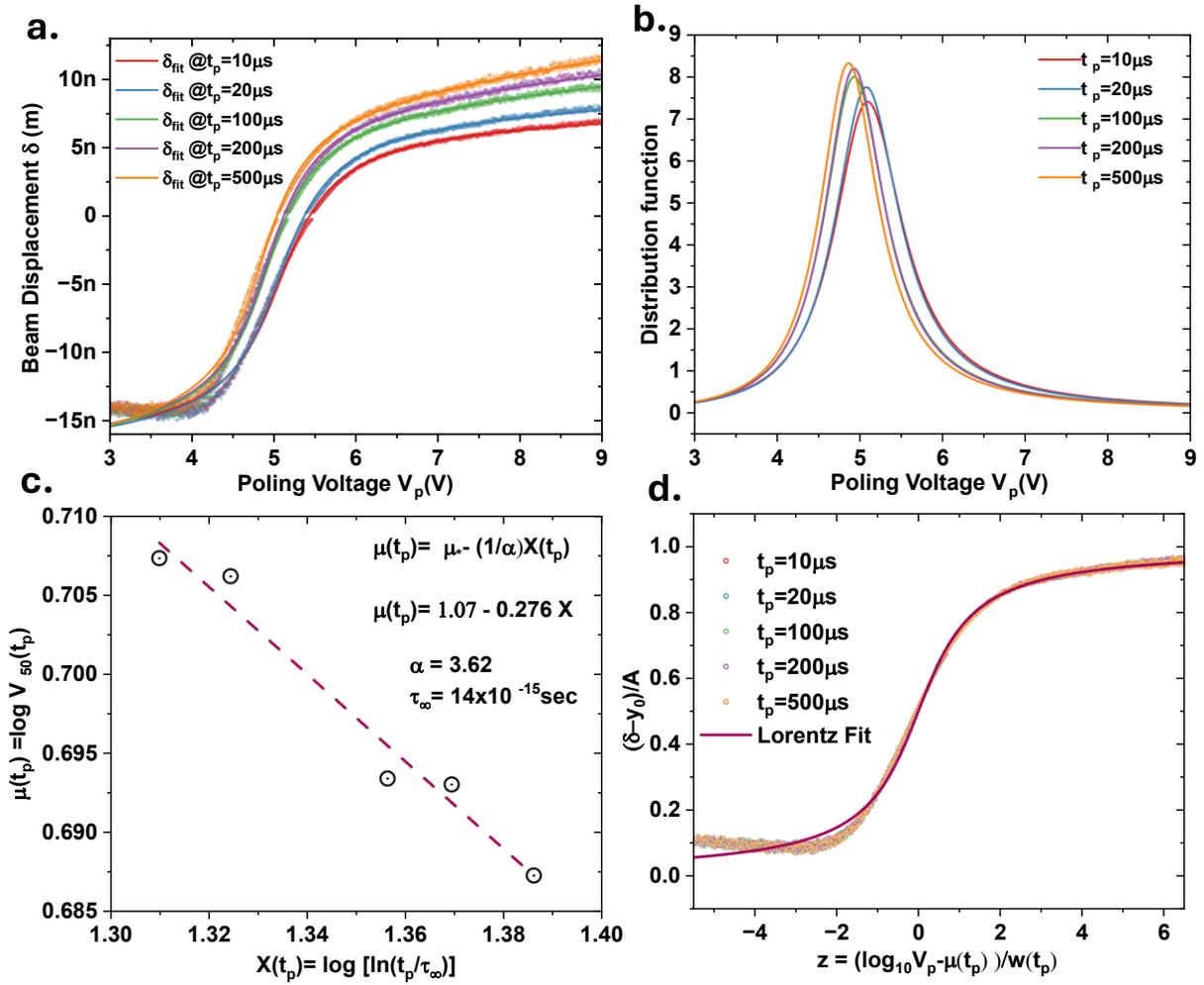

**Figure 4a.** Beam displacement $\delta(V_p)$ for multiple pulse widths $t_p = \{10, 20, 100, 200, 500\}\,\mu s$ at at fixed $V_{ac} = 250\,mV$. Points are data; solid curves are fits to a Lorentzian CDF in $\log V_p$. With increasing $t_p$, the transition shifts to lower $V_p$ (decreasing $\mu(t_p)$), consistent with field–time switching kinetics; $w(t_p)$ captures the breadth of the threshold distribution. **4b.** Distribution functions corresponding to Fig. 4a, obtained as the Lorentzian PDFs on $x = \log V_p$. Peak positions follow $\mu(t_p)$, widths reflect $w(t_p)$. **4c.** Rescaled collapse using the fit parameters from Fig. 4a. The normalized displacement $(\delta - y_0)/A$ is plotted versus the reduced coordinate $z = (\log V_p - \mu(t_p))/w(t_p)$. All datasets fall on a single parameter-free master curve, consistent with one underlying threshold distribution. **4d.** Merz-law regression of the median threshold. The fitted $\mu(t_p)$ values from Fig. 4a are plotted against $X = \log[\ln(t_p/\tau_\infty)]$. A linear fit yields $\alpha \approx 3.62$ and $\tau_\infty \approx 14 \times 10^{-15}$s; The negative slope shows that increasing $t_p$ lowers the effective threshold voltage, consistent with Merz-type field–time scaling.

The ability to program a large number of analog weight levels hinges on precisely determining the $\delta(V_p)$ transfer function. Our Lorentzian fits yield a full-width distribution of thresholds spanning roughly one decade in voltage. With a 18-bit digital-to-analog converter controlling $V_p$, we can resolve more than 256 distinct levels across the dynamic range. Since the mechanical readout is linear and effectively noise-free, level uncertainty is set mainly by stochastic domain switching. Across repeated writes, the standard deviation of $\delta$ at a fixed $V_p$ stays below 1 nm. (see Fig 3c). The median switching threshold $\mu(t_p)$ extracted from the Lorentzian fits obeys a Merz law. In the original empirical formulation, the switching time constant scales as $\tau \propto \exp(E_a/E)$, where $E$ is the electric field and $E_a$ is the activation field. For a triangular pulse of width $t_p$, the effective field is proportional to $V_p/t_p$, so equating the pulse duration to $\tau$ yields

$$(5)\ \log_{10} V_p = \mu(t_p) = \mu_* - \frac{1}{\alpha} X(t_p),$$

where $\tau_\infty$ is the intrinsic switching time in the infinite-field limit. Figure 4d plots $\mu(t_p)$ versus $\log[\ln(t_p/\tau_\infty)]$ for our data; the relationship is linear with slope $-1/\alpha$, yielding an activation field $\alpha = 3.62$ and $\tau_\infty = 14 \times 10^{-15}$ s. These values are comparable to those reported for PZT and HZO films and underscore that the same switching physics governs both electrical polarization and mechanical piezoelectric response. The width $w$ shows little dependence on $t_p$ over the range studied, implying that the distribution of pinning strengths is static and does not broaden or narrow significantly with pulse width.

Table 1. A comparison of Merz's law field dependence α from various works

| Authors | Merz's law field dependence α |
| --- | --- |
| Xiang et al. [41] | 3.602 |
| Alessandri et al.[42] | 3.73 |
| This work | 3.62 |

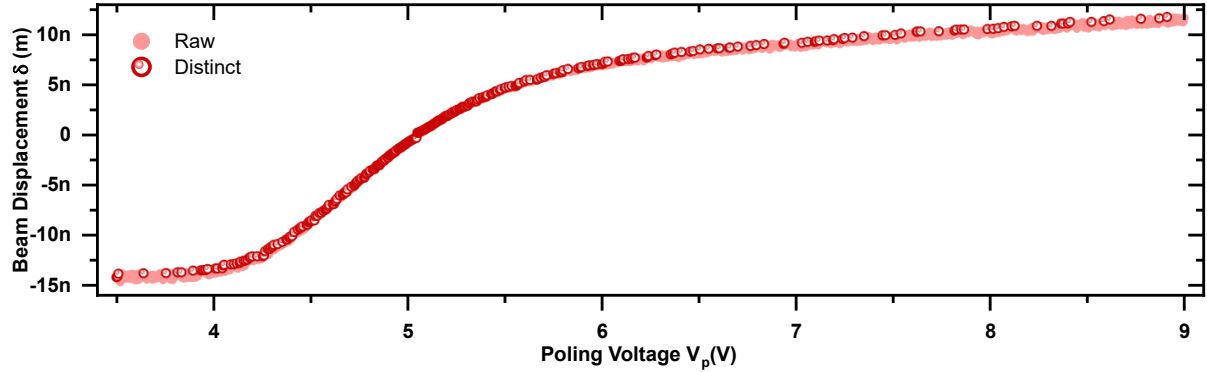

**Figure 5.** Graph showing distinct ~ 200 levels (> 7 bit) superimposed on the raw beam displacement data, obtained by following the strict monotonic subsequence scheme.

Accurate modelling of the weight transfer function enables calibration of intermediate levels and informs the design of programming protocols for neuromorphic systems.

Because the Lorentzian distribution is heavy-tailed, the fraction of domains switching at each incremental increase in $V_p$ decreases gradually; thus the weight update curve is smooth and highly reproducible. By targeting specific values of the reduced coordinate $z = (\log_{10} V_p - \mu)/w$, one can generate evenly spaced weight levels across the entire range. The Merz scaling further allows prediction of how the transfer function shifts with pulse duration, enabling dynamic adjustment of programming parameters to compensate for device-to-device variations or ageing. Together, the NLS–Lorentzian model and Merz law provide a compact analytical description of our mechanical synapse, facilitating integration into circuit-level simulations and neuromorphic learning algorithms. These insights build on earlier demonstrations of ferroelectric NEMS multipliers[36,37,43], extending them to the neuromorphic regime and highlighting the advantages of volume-integrated weight storage over conventional ferroelectric synapses. Finally, we demonstrate that our device can store high-precision weights suitable for neuromorphic hardware requiring 6–8 bit resolution following the previously mentioned write-read electro-mechanical scheme. We measure $\delta$ vs. $V_p$ while partially poling HZO and extract discrete levels by taking a strict monotonic subsequence as explained in S6. The number of distinct levels we obtained by this method was found to be ~ 200; the method transparently reveals the data's monotonic structure, enabling fine, stable weight quantization (see Figure 5).

We have demonstrated a neuromorphic weight element that stores analog values in the volume-integrated piezoelectric coefficient of a released HZO unimorph beam. Partial switching of ferroelectric domains tunes $d_{31}$, and a small-amplitude mechanical drive reads out the weight non-destructively. The mechanical weight transfer function follows a Lorentzian distribution in the common-logarithm ($\log_{10}$) of the programming voltage, consistent with a nucleation-limited switching model, and the median switching threshold obeys a Merz-type field–time scaling. These relationships allow us to calibrate and predict the mapping from programming parameters to weight values with high precision, achieving repeatable and reversible updates over many cycles. Compared with conventional ferroelectric synapses, volume-integrated piezoelectric coefficient $d_{31,eff}$ storage offers direct reliable read-out, negligible leakage, access to positive and negative weights, and, crucially, the ability to measure the fractional domain populations rather than just the net polarization. By leveraging the heavy-tailed distribution of switching thresholds and the universality of the Lorentzian scaling, we demonstrate the potential for high-bit-depth mechanical synapses suitable for neuromorphic inference and learning.

## Acknowledgement

This work was performed in part at the Cornell NanoScale Facility, a member of the National Nanotechnology Coordinated Infrastructure (NNCI), which is supported by the National Science Foundation (Grant NNCI-2025233). This work was supported by the DARPA NaPSAC (Award #: N660012424001)

# Supplementary Information

This document provides detailed derivations and background for the modelling used in the main manuscript. Equations are numbered with an "S" prefix for reference. Throughout we use the small-signal tip displacement $\delta$ as a linear proxy for the effective piezoelectric coefficient $d_{31,eff}$, the programming pulse voltage $V_p$ and width $t_p$, and the read voltage $V_{ac}$.

## S1. Domain-switching framework

### S1.1 Nucleation-limited switching and the switched fraction

Ferroelectric polarization reversal in polycrystalline films occur via Nucleation-Limited switching (NLS) mechanism and the switching time of the domain depends on localized electric filed E, given by Merz's law $\tau \propto \exp(E_a/E)$, where $E_a$ is the activation field and $E$ is the local electric field. In the NLS framework each $j^{th}$ grain (hysteron) switches independently with a characteristic switching time $\tau_j$. The ensemble-averaged switched fraction after a pulse of duration $t$ is obtained by integrating over a distribution $F(\ln \tau)$ of logarithmic switching times:

$$(S1)\ S(V, t) = \int_{-\infty}^{\infty} \left[1 - \exp\left(-\left(\frac{t}{\tau(V)}\right)^n\right)\right] F(\ln \tau(V)) \mathrm{d}(\ln \tau(V))$$

where $n$ is an effective dimensionality and $F$ encodes the broad dispersion in switching kinetics. When $F$ is narrow, Equation (S1) reduces to the classical Kolmogorov–Avrami–Ishibashi (KAI) model; when $F$ is broad, the switched fraction exhibits long tails even for long pulses, consistent with experiments on hafnia-based ferroelectrics[1–3].

Because the small-signal displacement is proportional to the piezoelectric coefficient and hence to the switched fraction, we relate $\delta$ to $S$ via an offset–span form

$$(S2)\ \delta(V_p, t_p) = \delta_{\min} + A\, S(V_p, t_p),$$

where $\delta_{\min}$ and $\delta_{\max}$ are the displacements of the fully reset and fully poled states, respectively. This relation defines the device-level offsets $\delta_{\min}$ and $A = K_{geom} V_{ac}(\delta_{\max} - \delta_{\min})$, which are held constant across pulse widths in the fitting procedure.

### S1.2 Detailed derivation of the Merz-law regression

The linear relation used to fit the medians in Fig. 4d follows directly from the generalized Merz law rather than being an empirical approximation[4]. Here we derive this relation step by step. In the generalized Merz law the switching time of a domain depends on the local electric field $E$ as

$$(S3)\ \tau(E) = \tau_\infty \exp\left[\left(\frac{E_a}{E}\right)^\alpha\right]$$

where $\tau_\infty$ is the attempt time, $E_a$ the activation field and $\alpha$ a dimensionless exponent. To find the programming field required to switch a domain within a pulse of duration $t_p$ we set $\tau(E^*) = t_p$ and invert Equation

$$(S4)\ E^*(t_p) = \frac{E_a}{[\ln(t_p/\tau_\infty)]^{1/\alpha}}$$

The corresponding threshold voltage is obtained by multiplying the threshold field by the thickness of the ferroelectric layer, $t_{HZO}$:

$$(S5)\ V_{50}(t_p) = E^*(t_p) \cdot t_{HZO} = \frac{E_a \cdot t_{HZO}}{[\ln(t_p/\tau_\infty)]^{1/\alpha}}$$

Where, $V_{50}(t_p)$ is the voltage at which switched fraction is 0.5.

Taking the base-10 logarithm defines the median of the logarithmic threshold distribution, $\mu(t_p) \equiv \log_{10} V_{50}(t_p)$. Separating terms and introducing $\mu_* = \log_{10}(E_a t_{HZO})$ and $X(t_p) = \log_{10}[\ln(t_p/\tau_\infty)]$ yields

$$(S6)\ \mu(t_p) = \mu_* - \frac{1}{\alpha} X(t_p),$$

Equation (S11) is the linear form used in the main text (Fig. 4d) to extract the dimensionless exponent $\alpha$ and the attempt time $\tau_\infty$. The slope of $\mu(t_p)$ versus $X(t_p)$ is $-1/\alpha$ and the intercept $\mu_*$ relates to the activation field via $E_a t_{HZO} = 10^{\mu_*}$. Once $\alpha$ and $E_a$ are known, $\tau_\infty$ follows from $\tau_\infty = t_p \exp[-(E_a/E)^\alpha]$ when evaluated at $E = V_{50}/t_{HZO}$. In practice, however, $\tau_\infty$ is determined more robustly by fitting the data across multiple pulse widths as described in S4.

## S2. Heavy-tailed disorder and the Lorentzian distribution

### S2.1 Origin of Lorentzian switching statistics

In polycrystalline ferroelectric films the local electric field experienced by each grain can differ from the applied field due to variations in grain geometry, dipole defects behaving as domain pinning sites, internal bias and interface asymmetry. These multiplicative variations in local field translate into additive fluctuations in the logarithm of the threshold voltage. A heavy-tailed distribution of these fluctuations leads naturally to a Lorentzian (Cauchy) distribution of switching thresholds. Such Lorentzian statistics have been observed to reproduce the entire polarization-switching kinetics over wide field ranges.

### S2.2 Cauchy cumulative distribution and displacement model

Following this heavy-tailed picture, we write the cumulative distribution function (CDF) for the switched fraction on the logarithmic voltage axis $x = \log_{10} V_p$ as a Cauchy distribution:

$$(S7)\ S(V_p, t_p) = \frac{1}{2} + \frac{1}{\pi}\arctan\left(\frac{\log_{10} V_p - \mu(t_p)}{w(t_p)}\right),$$

where $\mu(t_p)$ is the median and $w(t_p)$ the half-width of the distribution. Differentiating the CDF yields the Cauchy probability density function (PDF) on the log-voltage axis

$$(S8)\ f_{t_p}(x) = \frac{1}{\pi}\frac{w(t_p)}{[x - \mu(t_p)]^2 + w(t_p)^2}.$$

Substituting Equation (S7) into the offset–span relation (S2) provides a minimal, monotonic model for the displacement:

$$(S9)\ \delta(V_p, t_p) = y_0 + A\left[\frac{1}{2} + \frac{1}{\pi}\arctan\left(\frac{x - \mu(t_p)}{w(t_p)}\right)\right],$$

where $y_0 \equiv \delta_{\min}$ and $A \equiv K_{geam}V_{ac}(\delta_{\max} - \delta_{\min})$ are global offsets determined once per device. Equation (S9) is strictly monotonic in $V_p$; its derivative with respect to $x = \log_{10} V_p$ is the PDF in Equation (S8), reflecting the heavy-tailed distribution of local thresholds.

## S3. Fitting parameters and parameter table

The parameters $y_0$, $A$, $\mu(t_p)$ and $w(t_p)$ were extracted from the displacement curves $\delta(V_p, t_p)$ at each pulse width by nonlinear least-squares fitting of Equation (S9). Table S1 summarizes the values obtained from our experiments; the median $\mu(t_p)$ is quoted both in logarithmic units and as the physical half-switching voltage $V_{50} = 10^{\mu(t_p)}$, where $S(10^\mu) = 0.5$. The half-width $w(t_p)$ is in decades of $\log_{10} V_p$. The quantities $y_0$ and $A$ are held constant across pulse widths.

Table S1. Fitting parameters for different pulse widths

| Pulse width $t_p$ (µs) | Offset $y_0$ (nm) | Span $A$ (nm) | Median $\mu(t_p)$ | Half-width $w(t_p)$ | Threshold voltage $V_{50}$ (V) | Mechanical Coercive Voltage $V_{c,Mech}$ |
|---|---|---|---|---|---|---|
| 10 | −17.0472 | 23.7906 | 0.707319 | 0.042982 | 5.097 | 5.47 |
| 20 | −17.5778 | 24.2118 | 0.706183 | 0.041078 | 5.084 | 5.395 |
| 100 | −18.2397 | 24.3556 | 0.693400 | 0.039788 | 4.936 | 5.175 |
| 200 | −18.4336 | 24.4328 | 0.693020 | 0.038840 | 4.932 | 5.135 |
| 500 | −19.1364 | 24.2516 | 0.687272 | 0.038244 | 4.867 | 5.05 |

*Note.* These values correspond to the curves shown in the main manuscript. The nearly constant half-width indicates that disorder dominates the breadth of thresholds, while the median decreases with pulse width, reflecting Merz-type kinetics.

## S4. Parameter extraction and Merz scaling

To extract the kinetic parameters $\alpha$ and $\tau_\infty$, we regress the medians $\mu(t_p)$ against $X = \log_{10}[\ln(t_p/\tau_\infty)]$. From the linear relation

$$(S10)\ \mu(t_p) = \mu_* - \frac{1}{\alpha} X(t_p),$$

the slope yields $1/\alpha$ and the intercept $\mu_*$ sets $\tau_\infty$ via $\mu_* = \log_{10}[E_a t_{HZO}]$. Fitting our data gives $\alpha = 3.62$ and $\tau_\infty \approx 14 \times 10^{-15}$ s. The negative slope confirms that longer pulses lower the effective switching threshold. These values are consistent with other reports of Merz-type scaling in hafnia-based ferroelectric thin films.

## S5 Reduced coordinates and universal collapse

For visualization and comparison across pulse widths we rescale the displacement data using the reduced coordinate $z = \left(\log_{10} V_p - \mu(t_p)\right)/w(t_p)$. The normalised displacement $(\delta - y_0)/A$ plotted versus $z$ collapses all datasets onto the parameter-free Cauchy function

$$(S11)\ S(z) = \frac{1}{2} + \frac{1}{\pi}\arctan z,$$

as shown in Fig. 4c. This collapse demonstrates that, after appropriate rescaling, the mechanical response follows the universal Lorentzian statistics expected for heavy-tailed disorder independent of pulse width. This collapse underscores that multiplicative disorder, rather than deterministic KAI dynamics, governs the domain switching statistics in our hafnia-based ferroelectric unimorph. The reduced coordinate also highlights the role of the median $\mu(t_p)$(which sets $z = 0$) and the width $w(t_p)$ in controlling the shape of the weight-transfer curve. In particular, $z = 0$ corresponds to $V_p = 10^{\mu(t_p)} \approx V_{c,mech}$ where the mechanical and electrical weights are balanced.

## S6 Extracting Distinct Levels with the Strict Monotonic (S0) Filter

In our study, we record $\delta$ as a function of the poling voltage $V_p$ during partial polarization of HZO films. As the true displacement–voltage response is monotonic, we convert this continuous set of measurements into discrete levels, which reflect the sequence of increasing displacements. The simplest way to achieve this is to construct a *strict monotonic subsequence* of the raw data. A function $f$ is *monotonically increasing* (also called non-decreasing) if for any two inputs $x$ and $y$ such that $x \leq y$, the outputs satisfy $f(x) \leq f(y)$. Our filter follows this definition directly.

### Algorithm

1. Sort the data by voltage. We begin with the raw measurements $\{(V_{p,i}, y_i)\}$ and sort them in ascending order of $V_{p,i}$. In practice the data are already acquired with increasing voltage, so no reordering is required.

2. Initialize the first level. We accept the first displacement value $y_1$ and call this the initial kept value, setting a variable $y_{last} \leftarrow y_1$ and assigning it level index $k = 1$.

3. Scan and keep non-decreasing values. For each subsequent index $i = 2, \ldots, N$, we compare the new measurement $y_i$ to the last kept value $y_{last}$. If

$$y_i \geq y_{last},$$

then we accept $y_i$ as a new level, increment the level index $k \leftarrow k + 1$, and update $y_{last} \leftarrow y_i$. Otherwise we discard $y_i$ as an overlap because it is smaller than a previously accepted point. This rule enforces the non-decreasing condition of a monotonic function: for larger voltages, the kept displacements never fall below earlier kept values[1].

4. Assign level indices. Each accepted sample is labelled by its integer level index $k$. The number of levels $K$ equals the number of accepted points.

## Distinct-Level Data

The output of the S0 filter is the sequence of kept points $(V_{p,k}, y_k)$ together with their level indices. Because the algorithm accepts a point whenever it does not decrease, it yields the maximum possible number of levels consistent with monotonicity. These levels correspond to the steps of a staircase function $L(V_p)$ defined as

$$L(V_p) = \text{number of kept points with voltage } \leq V_p.$$

This staircase increases by one whenever a new kept displacement is encountered and remains constant in regions where measurements fall below the last kept value. . Adjacent kept points are separated by variable displacements; where the response is nearly flat, the filter finds many closely spaced levels, while in regions with larger increments it finds fewer.

## Interpretation

The strict monotonic filter implements the textbook definition of a non-decreasing function, if a later point is lower than a previously accepted value, it is ignored. Because no explicit noise margin is used, the S0 levels are "upper bounds" on how finely the continuous curve could be digitized. In practice, experimental noise can cause very small positive fluctuations that produce new levels even though the underlying displacement has not changed. Nonetheless, the S0 filter provides a transparent way to visualize the monotonic structure of the raw data and to assign an integer level index for each voltage.